\documentclass[twocolumn,showpacs,superscriptaddress,pre]{revtex4}
\usepackage{latexsym}
\usepackage{graphicx}
\usepackage{amssymb}
\input{epsf}
\begin{document}
\title{A Map-Based Model of the Cardiac Action Potential}

\author{Nikolai F. Rulkov}
\affiliation{Institute for Nonlinear Science, University of
California, San Diego, La Jolla, CA 92093-0402,}
\affiliation{Information Systems Laboratories, Inc., San Diego, CA
92121}

\date{\today}
\begin{abstract}
A discrete time model that is capable of replicating the basic
features of cardiac cell action potentials is suggested. The paper
shows how the map-based approaches can be used to design highly
efficient computational models (algorithms) that enable large-scale
simulations and analysis of discrete network models of cardiac
activity.

\end{abstract}

 \maketitle

\vskip1pc
\newcounter{eq}

\section{Introduction}

\begin{figure}[b]
\includegraphics[scale=0.4]{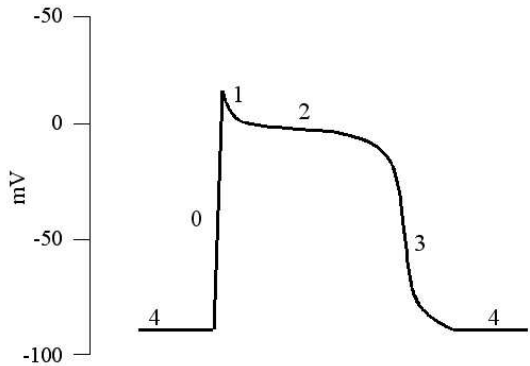}
\caption{ A sketch of typical action potential of a ventricular cell
illustrates the main phases associated with different states of
channel activity. }\label{fig-cell-AP}
\end{figure}

The heart cells that are directly involved in the dynamics of its
electrical activity include pacemaker and non-pacemaker cells. The
peacemaker cells generate spontaneous action potentials and are
characterized by a slow rate of the depolarization. These cells are
found in sinoatrial and atrioventricular nodes of the heart and
initiate the propagation of electrical activity throughout the
heart. The non-pacemaker cells, involved in the propagation of
electrical activity, such as atrial myocytes, ventricular myocytes
and Purkinje cells, have a specific form of action potential (AP)
which is characterized by five main phases including a very rapid
depolarization and a prolonged plateau~\cite{Eick81}, see
Fig~\ref{fig-cell-AP}. This paper proposes a simple computationally
efficient model that can replicate these phases of AP.

The non-pacemaker cell has a very negative resting potential (phase
4) characterized by the open potassium channels and, therefore,
K$^+$ current and the closed fast sodium Na$^+$ and slow calcium
channels Ca$^{2+}$. Being depolarized to the threshold voltage of
about -70mV the cell shows a very rapid depolarization caused by
opening fast sodium channels Na$^+$ (phase 0). This phase quickly
increases the membrane potential to positive values where the Na$^+$
channels become inactivated and the dynamics of the membrane changes
to an initial repolarization (phase 1) induced by a short-term
transient outward K$^+$ current. The plateau of cardiac action
potential (phase 2) is the result of balanced dynamics between
inward Ca$^{2+}$ current and the outward K$^+$ current coming
through the slow delayed rectifier potassium channels. A number of
other ionic currents are also involved in this phase. As the
Ca$^{2+}$ channels start to close while the rectifier K$^+$ channels
remain open, the membrane potential drops to the levels of resting
potential forming phase 3. Even from this, an overly simplified
scenario involved in the formation cardiac action potential, it
becomes clear that any attempt to produce a conductance based model
of a non-pacemaker cell will lead to a large system of differential
equations. A number of such models have been proposed and are used
as accurate models of cardiac cells, see for
example~\cite{LuoRudy94a,LuoRudy94b,NobleRudy01}

The simplified approaches to the modeling of cardiac ventricular
action potentials include the sets of ODE models where each ODE
equation phenomenologically represents the dynamics of multiple
channels (for example, van der Pol equation~\cite{vanderPoll28} or
the FitzHugh-Negumo model~\cite{FitzHugh61,Rogers94}). However, due
to the high depolarizing rates of AP during phase 0 numerical
simulations with the ODE models require a significant reduction of
the integration step size which complicates the use of these simple
models in studies of large scale networks. Here we suggest a model
of the cardiac AP which is built using a discrete time map and
designed to significantly increase the time step of simulations. A
similar approach was successfully used before in the design of
computationally efficient models of
spiking-bursting~\cite{Rulkov02}, regular spiking and fast and
spiking neurons~\cite{RTB_CN04}, and other neurons.

\section{Map-based model of cardiac AP}\label{sec1}

The simplest form of the suggested model is a two-dimensional map,
which can be written as \setcounter{eq}{\value{equation}}
\addtocounter{eq}{+1} \setcounter{equation}{0}
\renewcommand{\theequation}{\theeq \alph{equation}}

\begin{eqnarray}
 x_{n+1}&=&P(x_n,y_n)~, \label{mapx} \\
 y_{n+1}&=&Q(x_n,y_n)~, \label{mapy}
\end{eqnarray}
\newcommand{\refmap}{1}
where the dynamics of $x_n$ represents the fast changes related the
phase 0 and the dynamics of $y_n$ forms the action potential during
the remaining phases. The voltage of action potential $V_n$ will be
defined at the end of this section as a liner combination of $x_n$
and $y_n$.

\setcounter{equation}{\value{eq}}
\renewcommand{\theequation}{\arabic{equation}}

\begin{figure}[t]
\includegraphics[scale=0.4]{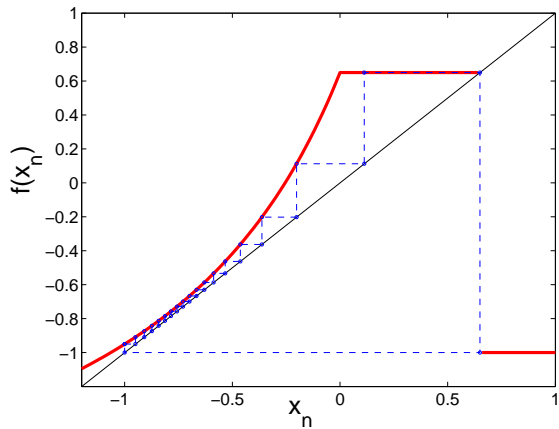}
\caption{(Color online) The shape of $f(x_n,u)$ and a limit cycle
generated by subsystem (\ref{mapx}) with $\varepsilon(y_n)=0$ and a
fixed value of $u$=-2.55 }\label{fig2}
\end{figure}

The function $P(x_n,y_n)$ in the right hand side of the first
equation (\ref{mapx}) can be written as
\begin{eqnarray}
P(x_n,y_n)=(1-\varepsilon(y_n))f(x_n,x_{n-1},u)+\varepsilon(y_n)x_p,
 \label{P}
\end{eqnarray}
where the nonlinear function $f(x_n,x_{n-1},u)$ is of the
form~\cite{Rulkov02}
\begin{eqnarray}
&f(x_n,x_{n-1},u)=  \label{func} \\
 & \cases{
\alpha/(1-x_n)+u,  & if $x_n \leq 0$ \cr \alpha+u,          & if $0
< x_n<\alpha+u $ and $ x_{n-1}\leq0$ \cr -1,                & if
$x_n \geq \alpha+u $ or $ x_{n-1}>0.$ \cr} \nonumber
\end{eqnarray}
Note, that for a description of autonomous dynamics of the model the
conditions related to the values of $x_{n-1}$ can be omitted. The
third argument $u$ will represent a linear combination of the
function parameter $\beta_x$ and the input variable (e.g. injected
current)
\begin{eqnarray}
u=\beta_x + I_n. \label{u}
\end{eqnarray}
Parameter $\alpha$ is a control parameter of the map. The dependence
of $f(x,u)$ on $x$ computed for fixed values of $u$ and
$\varepsilon(y_n)=0$, i.e. when $P(x_n,y_n)=f(x,u)$, is shown in
Fig.\ref{fig2}. This figure also illustrates a trajectory of an
uncoupled one-dimensional map (\ref{mapx}). The limit cycle of the
map was used in~\cite{Rulkov02} to replicate a sequence of short
neuronal action potentials - a spike train. Note, that the third
condition of $f(x,u)$ corresponds to the moment of time when $x_n$
reaches its maximum value, i.e. the tip of a spike.

To replicate the action potential of a cardiac ventricular cell we
shift $f(x,u)$ down where the nonlinear segment intersects the
diagonal giving birth to stable and unstable fixed points. This
shift destroys the limit cycle shown in Fig.\ref{fig2}. In addition
to that we modulate the 1-D map with $\varepsilon(y_n)$, where
$0\leq \varepsilon(y_n) \leq 1$, see (\ref{P}). One can see from
(\ref{P}) that when $\varepsilon(y_n)$ approaches one the $x$
dependance of function $P(x_n,y_n)$ flattens and approaches the
values $P(x_n,y_n) \rightarrow x_p$ for all $x_n$. Therefore, the
increase of $\varepsilon(y_n)$ "deactivates" the motions of the 1-D
map (\ref{mapx}) by forming a superstable fixed point $x_p$. We will
utilize this effect to replicate the blocking of Na$^+$ channels to
terminate phase 0, see Fig~\ref{fig-cell-AP}.

\begin{figure}[t]
\includegraphics[scale=0.4]{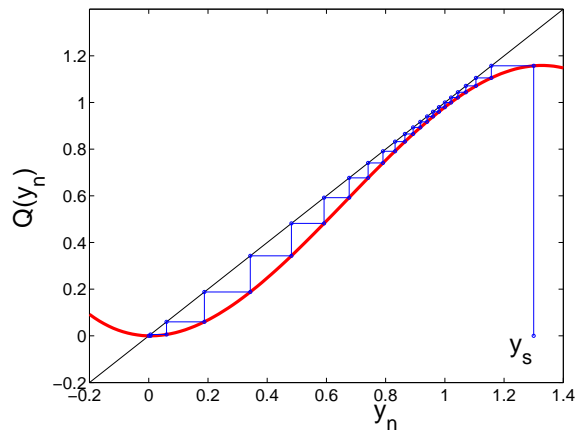}
\caption{(Color online) The dependance of $Q(x_n,y_n)$ on $y_n$ and
the trajectory of subsystem (\ref{mapy}) started at $y_s$=1.3
plotted for the parameter values $\mu=0.02$ and $g=1.0$.
}\label{funQ}
\end{figure}

\begin{figure}[h]
\includegraphics[scale=0.35]{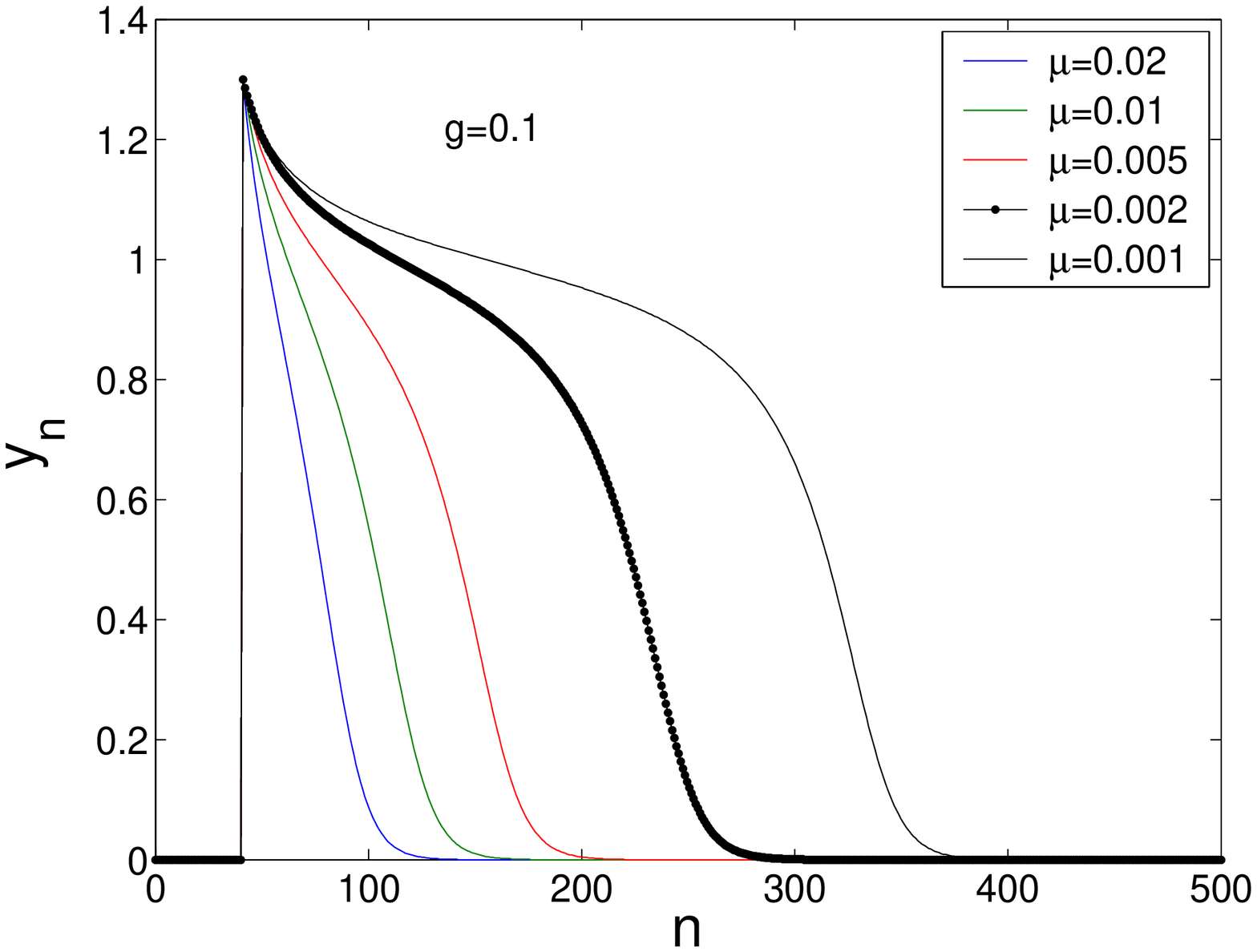}
\includegraphics[scale=0.35]{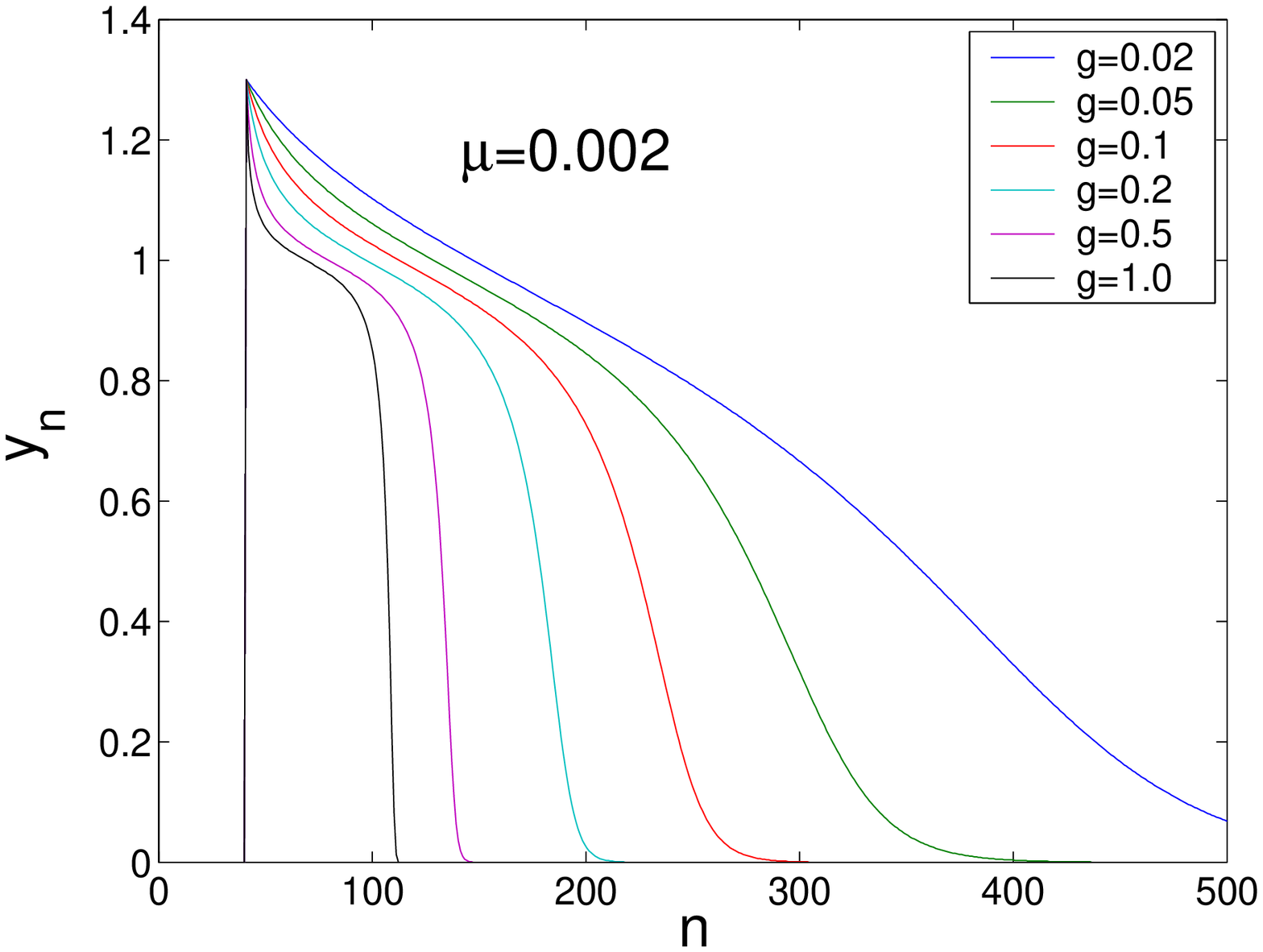}
\caption{(Color online) Waveforms generated by subsystem
(\ref{mapy}) with $y_s$=1.3 and different values of parameters $\mu$
(top panel) and $g$ (bottom panel). }\label{wfQ}
\end{figure}

When equation (\ref{mapx}) produces a spike the membrane potential
rises and the other ionic channels get involved the formation of the
plateau described above as phases 1,2 and 3. The dynamics of the
cell during this part of AP is governed by subsystem (\ref{mapy}),
where the right hand side can be written in the form
\begin{eqnarray}
&Q(x_n,y_n) =  &\cases{y_s,     &  if $x_n \geq \alpha+u $ or $
x_{n-1}>0,$ \cr q(y_n),     &  otherwise. \cr}\ \label{Q}
\end{eqnarray}
where $$q(y_n)=(1-\mu) y_n-gy_n(1-y_n)^2.$$ Note that the first
condition of (\ref{Q}) is the same as the third condition of
(\ref{func}) and corresponds to the moment of spike in subsystem
(\ref{mapx}). One can see that at the moment of spike the trajectory
of the subsystem (\ref{mapy}) starts at the value $y_n=y_s$ and then
follows the dynamics of the one-dimensional map which is modeled
here with a polynomial function $q(y_n)$. Function $q(y_n)$ is
designed to form a stable fixed point at $y_n$=0 and a very narrow
gap between the function and the diagonal where the trajectories of
subsystem (\ref{mapy}) slow down, see Fig.~\ref{funQ}. This slow
motion is used to form the plateau in the shape of cardiac action
potential, i.e. phase 2. The size of the gap and, therefore, the
duration of the plateau, is set by selecting parameter $\mu$. The
other control parameter $g$ of the function is used to shape the
action potential at the transient from plateau to the resting state,
i.e. phase 3. The idea behind the selection of these parameter
values is illustrated in Fig~\ref{wfQ}. The duration and shape of
the action potential can also be controlled by the location of
starting point in equation (\ref{mapy}), i.e. parameter $y_s$.

In order to model the absolute refractory period (ARP) that
prohibits triggering a new action potential during phases 1, 2 and
the beginning part of phase 3 we deactivate subsystem (\ref{mapx})
using function $\varepsilon(y_n)$ which can be defined as a step
function
\begin{eqnarray}
&\varepsilon(y_n) = &\cases{0,     &  if $y_n \leq y_{th}$ \cr 1 , &
if $y_n > y_{th}$. \cr}\label{eps}
\end{eqnarray}
Here, the value of $y_{th}$ sets the threshold of deactivation.

Equation (\ref{eps}) closes the feedback between subsystems
(\ref{mapx}) and (\ref{mapy}), and completes the basic part of the
map-based model. The membrane potential replicated by this model can
be defined as
\begin{eqnarray}
V[mV] = -25 mV + x_n \times 80 mV + y_n \times 85 mV.\label{Vmv}
\end{eqnarray}

\begin{figure}[h]
\includegraphics[scale=0.35]{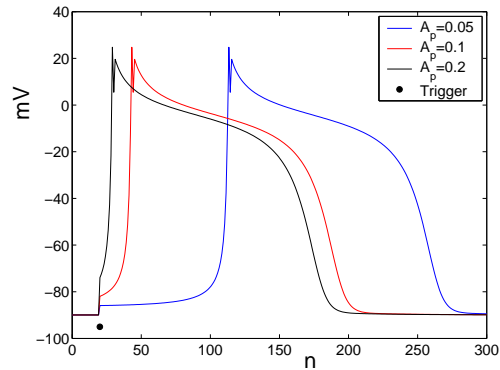}
\caption{(Color online) Example of ventricular action potentials
computed with formula (\ref{Vmv}). The parameter values are
$\alpha=3.2$, $\beta_x=-2.5780$, $x_p=-0.8$, $y_s=1.3$, $\mu=0.002$,
$g=0.2$, $y_{th}=0.5$ and three amplitudes $A_p$ of triggering
pulses of external current $I_n$ which is applied at $n=20$ for one
iteration. }\label{wfAP}
\end{figure}

Examples of APs produced by this model are presented in
Fig.\ref{wfAP}. The parameters $\alpha$ and $\beta_x$ of function
(\ref{func}), (\ref{u}) are set to provide a stable fixed point in
the subsystem (\ref{mapx}) at the negative values of $x_n$. This
takes place when the nonlinear part of $f(x_n)$ crosses the
diagonal. The AP was triggered by a pulse of external current $I_n$,
see (\ref{u}). This pulse moves function $f(x_n,u)$ up and, if the
amplitude of the pulse is sufficient, the stable fixed point
disappears via a saddle-node bifurcation in (\ref{mapx}). After that
the trajectory $x_n$ goes up forming a short spike in the waveform
of $x_n$. The time interval between the trigger pulse and the spike
depends on the amplitude and duration of the pulse, see
Fig.~\ref{wfAP}. The spike initiates the motion in subsystem
(\ref{mapy}) by changing its state to $y_n=y_s$, see (\ref{Q}).
After $y_n$ has occurred at the high level it deactivates subsystem
(\ref{mapx}) by setting $\varepsilon(y_n)=1$ and keeps it
deactivated until $y_n$ gets to the levels below $y_{th}$, see
(\ref{eps}).

\begin{figure}[h]
\includegraphics[scale=0.4]{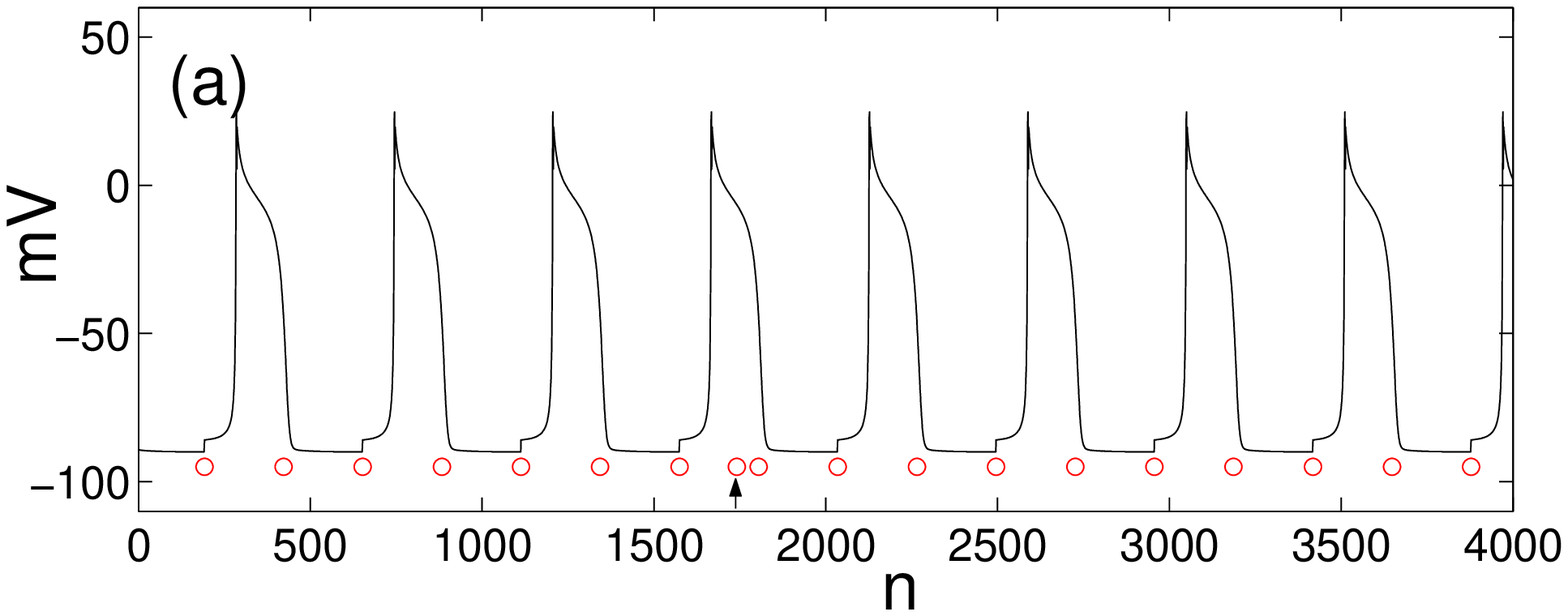}
\includegraphics[scale=0.4]{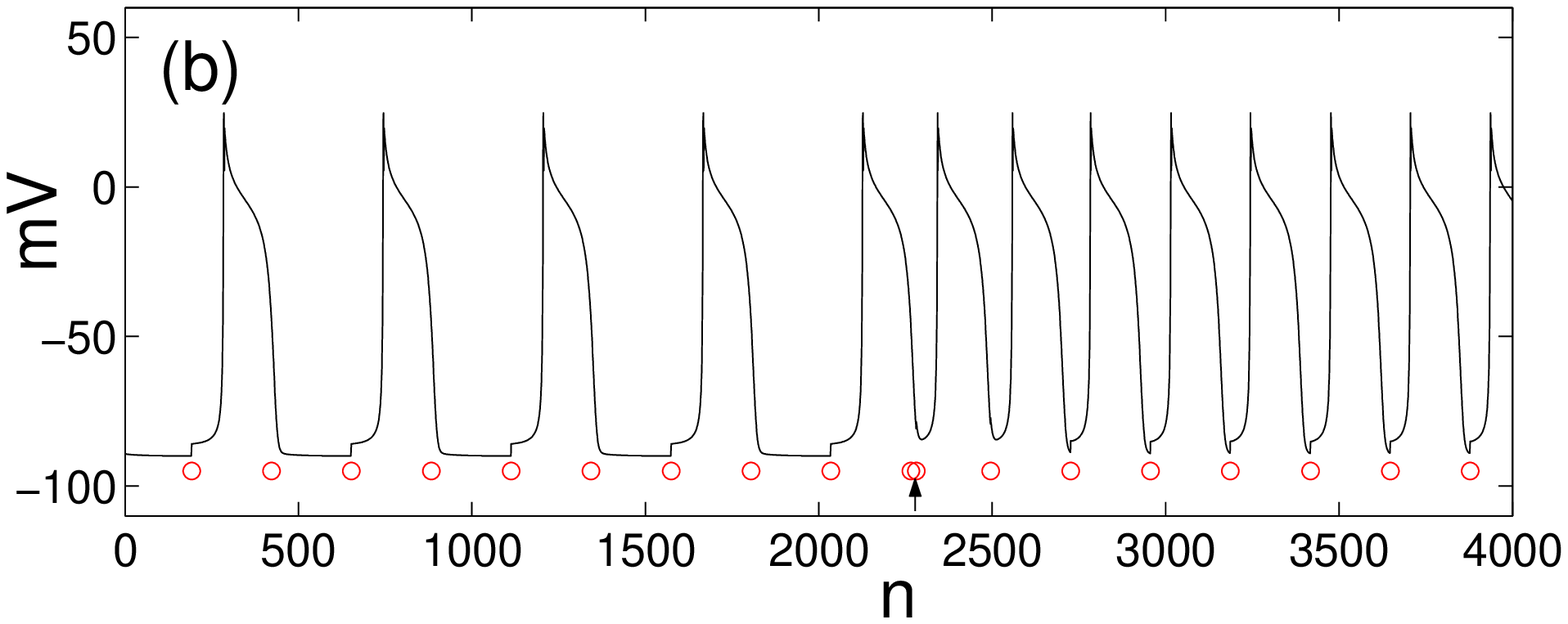}
\caption{(Color online) Action potentials triggered by a periodic
sequence of pulses (red circles) and one additional pulse indicated
by the arrow. The amplitude of pulses $A=0.05$. The other parameters
the same in Fig.~\ref{wfAP}. (a) Regime of oscillations with
frequency ratio 1:2. (b) Transition from 1:2 to 1:1 frequency ratio
caused by the additional pulse.}\label{wfpAP}
\end{figure}

Despite the simplicity of dynamical mechanisms included in the
map-based model constructed above the model is capable of
replicating a number of interesting properties of the behavior of
ventricular cells. For example it can capture the effects of the
coexistence of different oscillation regimes triggered by a periodic
sequence of pulses. With proper selection of frequency and amplitude
of the pulses the model can produce APs with the frequency ratio 1:2
or 1:1. These regimes are shown in Fig.~\ref{wfpAP}. When in
addition to the periodic sequence of pulses we add one more pulse
unrelated to the periodic sequence the result depends on the phase
of AP where the new pulse occurs. If the additional pulse falls
within the absolute refractory period it does not affect the
oscillations, see Fig.~\ref{wfpAP}a. If the pulse occurs before or
after the ARP it can switch the oscillations to the regime with a
different frequency locking ratio, as it is shown in
Fig.~\ref{wfpAP}b. The coexistence of these regimes indicate the
presence of a memory effect in the dynamics of cardiac AP. In this
basic model the memory is the result of a transient in subsystem
(\ref{mapx}) from the state $x_p$ to the fixed point after it has
been activated. Similar phenomena caused by a memory effect have
been studied in ~\cite{Hall99,Tolkacheva02,Schaefer07} with a
different type of mapping model.

\section{Modeling of memory effects}\label{sec3}

\begin{figure}[t]
\includegraphics[scale=0.4]{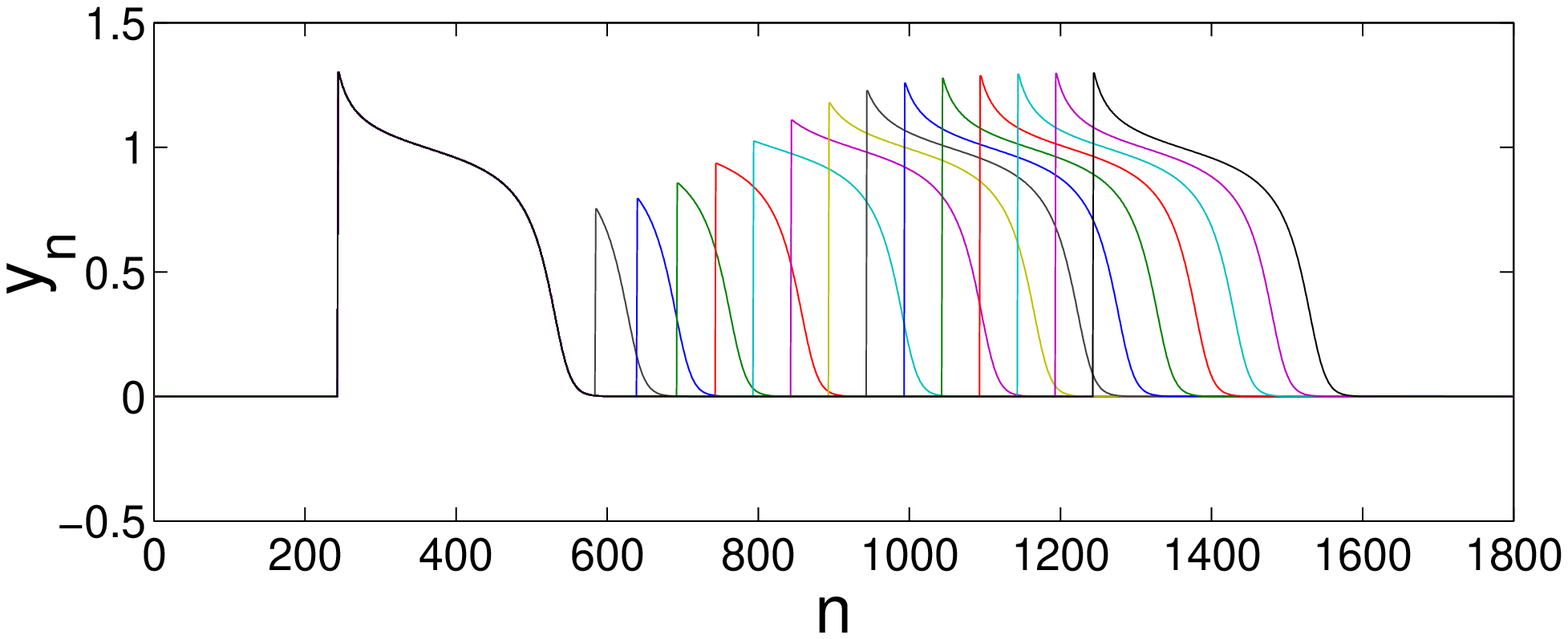}
\includegraphics[scale=0.4]{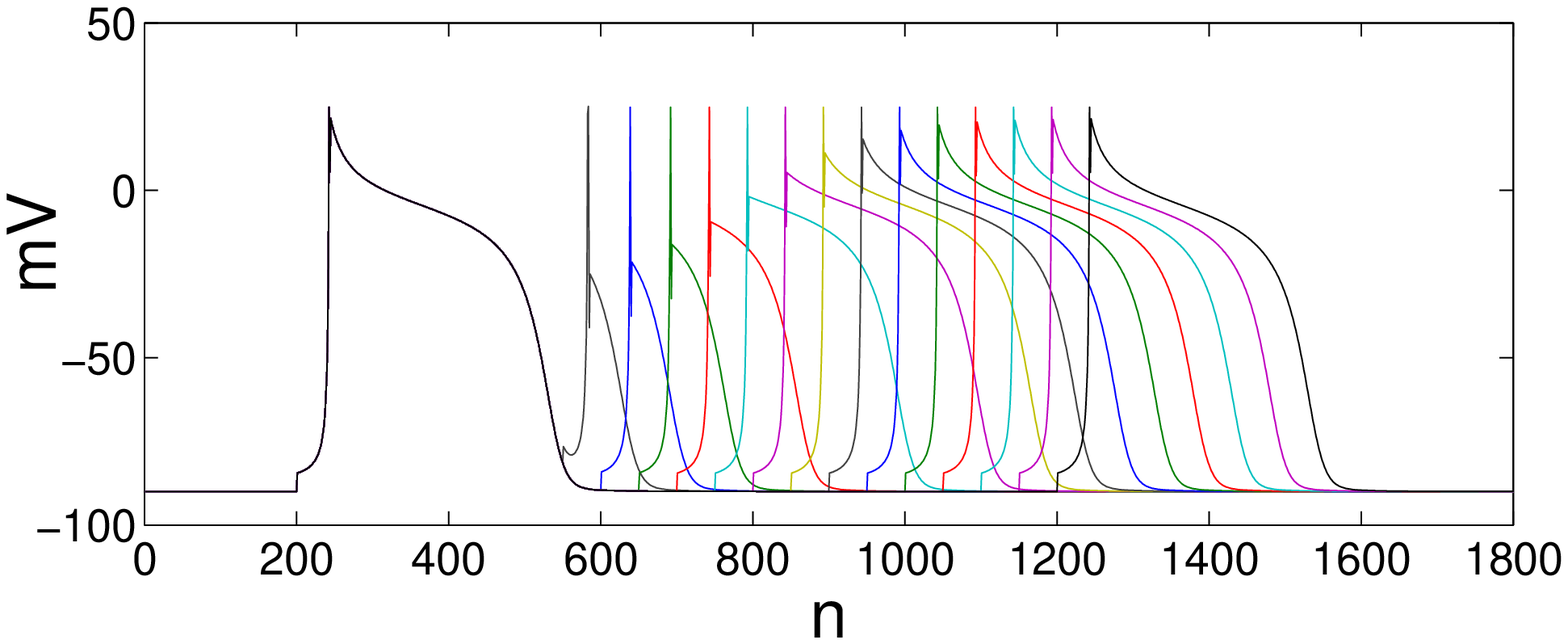}
\caption{(Color online) Waveforms of $y_n$ and AP illustrating the
effect of electrical restitution modeled by means of $y_s$
modulation using Eq.~\ref{Sn} with parameter values $S_a=0.6$,
$S_s=0.599$, $g_s=0.02$. The parameter values of the map-models are
the same as in Fig.~\ref{wfAP} except for $A=0.69$, $g=0.1$ and
$\mu=0.001$. }\label{wfAdapt}
\end{figure}

Regulation and rate dependance of action potential duration (APD) is
an important property of the ventricular
cells~\cite{LuoRudy94b,HundRudy04,Franz03}. In the considered
map-based model the effects related to APD adaptation can be
achieved by modulation of the subsystem (\ref{mapy}) by dynamically
varying the values of  parameters $\mu$, $g$ or $y_s$, see
(\ref{Q}). As an example consider the case when the parameter $y_s$
in (\ref{Q}) is substituted with a new variable $y_s^{new} = y_s
-s_n$ and $s_n$ is described by equation of the form
\begin{eqnarray}
s_{n+1} =  \cases{S_s,     &  if $x_n \geq \alpha+u $ or $
x_{n-1}>0,$ \cr q_s(s_n),     &  otherwise. \cr}\ \label{Sn}
\end{eqnarray}
where $q_s(s_n) = s_n-g_s s_n(S_a-s_n)$. This equation works similar
to subsystem (\ref{mapy}),(\ref{Q}). When subsystem (\ref{mapx})
generate a spike the trajectory of (\ref{Sn}) starts at $S_s$, ($0
\leq S_s< S_a$) and drifts to $s_n=0$ with the rate controlled by
parameter $g_s$, ($0<g_s<S_a^{-1}$). By selecting parameter values
$S_a=0.6$, $S_s=0.599$, $g_s=0.02$ one can tune the evolution of
$s_n$ to replicate the properties of the electrical restitution
curves. An example of such an adaptive behavior is shown in
Fig.~\ref{wfAdapt}. The figure presents a set of waveforms of two
consecutive action potentials. The first AP is triggered at n=200
while the time of the second AP is varied by changing the timing of
the second trigger pulse. One can see that due to dynamical
modulation of $y_s$ the duration of the second AP is significantly
reduced at the short time intervals between the trigger pulses. The
APD recovers as the time interval increases. The shape of the
restitution curve can be controlled by the parameters of system
(\ref{Sn}).

\section{Modeling of electric coupling}\label{sec4}

An important component in the simulations of electrical activity of
heart tissue is the model for electrical coupling among the cells.
This coupling occurs through a gap junction between the nearby
cells. A number for models describing the gap junction dynamics at
different levels of sophistication have been suggested and
reported~\cite{Joyner86,Jamaleddine96,Henriquez01}. Here we will
consider how the simplest gap junction model can be used to couple
the cells in the form of map-based models. We assume that current
flowing through the gap junction from cell $j$ into cell $i$ is
\begin{eqnarray}
I_{n,j,i}^{gap}= g_{gap} (V_{n,j}-V_{n,i})\label{Igap}
\end{eqnarray}
where $V_{n,i}$ is a membrane potential of cell $i$ given by Eq.
(\ref{Vmv}) and $g_{gap}$ is the conductance of the gap junction,
i.e. coupling strength parameter.

We will assume that at different phases of AP the current has
different effects on the dynamics of the cell. For example, at the
phases 0 and 4 the effect of coupling is more pronounced than during
the phases 1,2 and 3. Therefore, to insert the gap junction current
into the map-based model at the phases 0 and 4 we rewrite equation
(\ref{u}) by adding $I_{n,j,i}^{gap}$, i.e.
\begin{eqnarray}
u=\beta_x + I_n + \sum_{j\in J}(I_{n,j,i}^{gap}), \label{ugap}
\end{eqnarray}
where $J$ is the set of nearby cells. In order to capture the effect
of the gap junction during the phases 1,2 and 3, when subsystem
(\ref{mapx}) is turned off, one can rewrite equation (\ref{mapy}) as
follows
\begin{eqnarray}
 y_{n+1}&=&Q(x_n,y_n)+ \mu_{gap} \sum_{j\in J}(I_{n,j,i}^{gap})~, \label{mapygap}
\end{eqnarray}
where parameter $\mu_{gap}$ can be selected to set a proper balance
between the coupling strengths at different phases of AP.

\begin{figure}[t]
\includegraphics[scale=0.4]{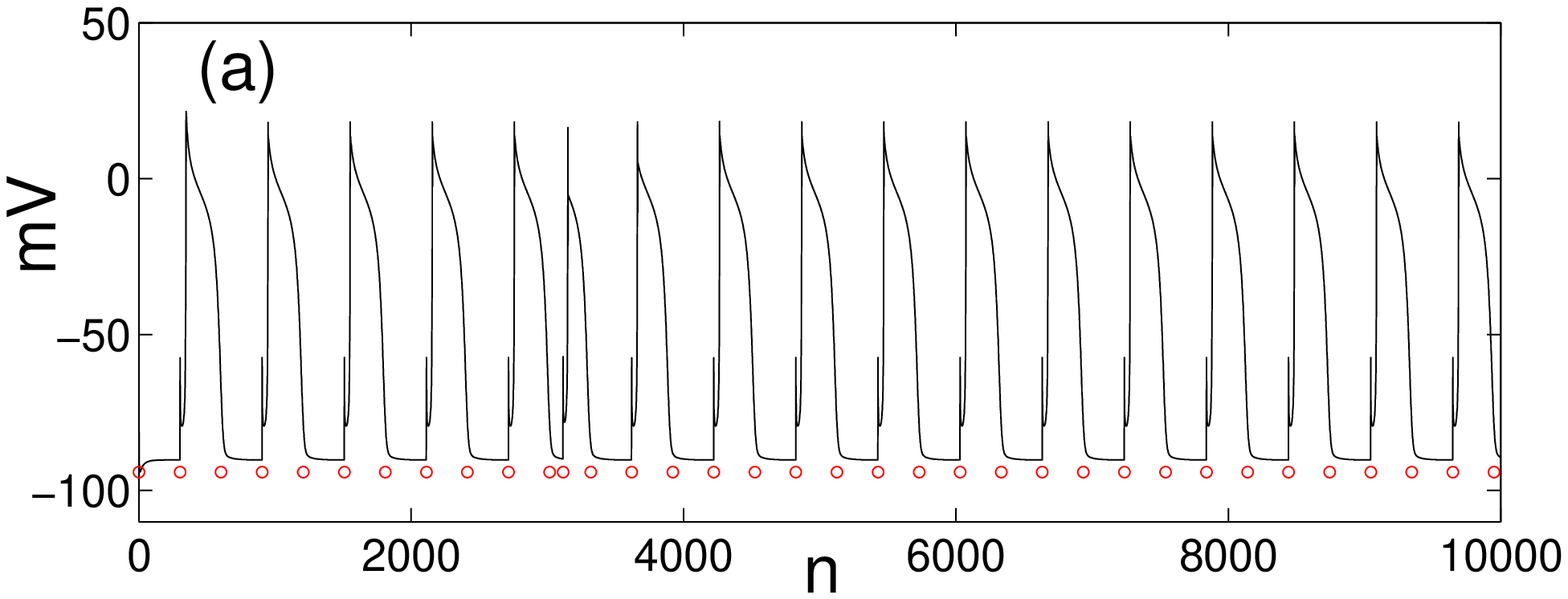}
\includegraphics[scale=0.4]{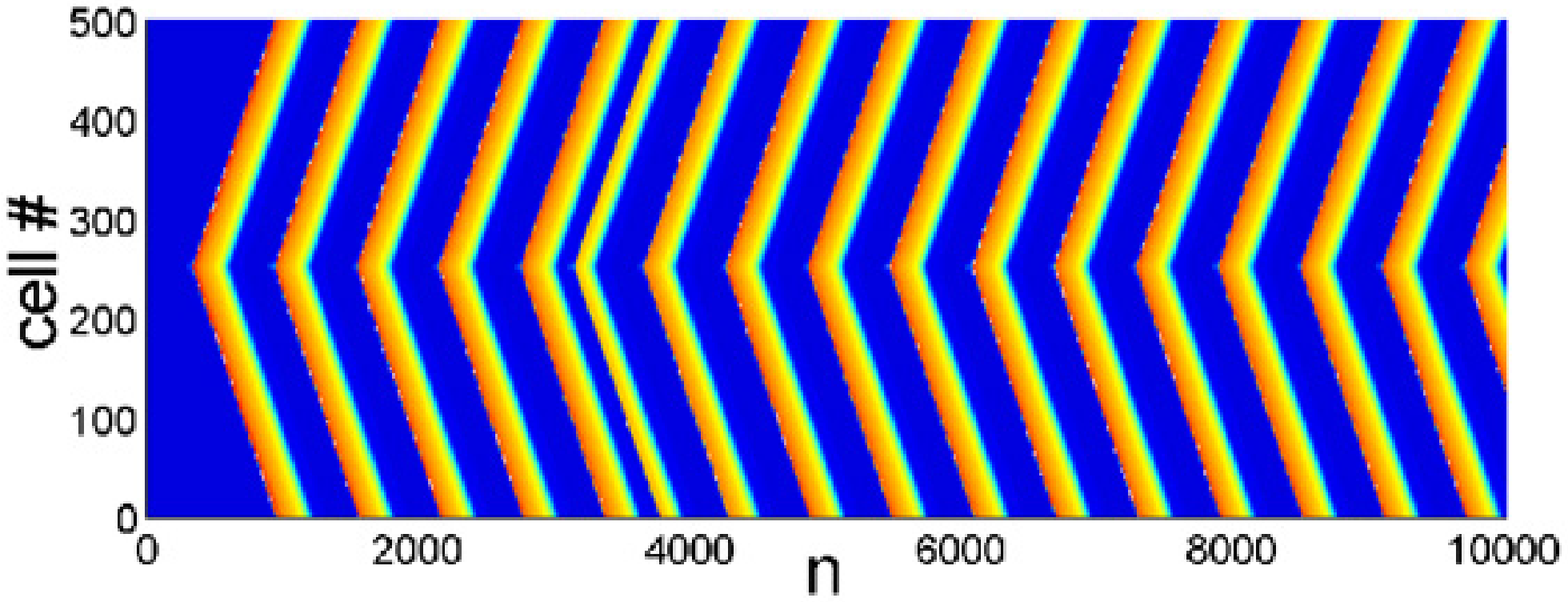}
\includegraphics[scale=0.4]{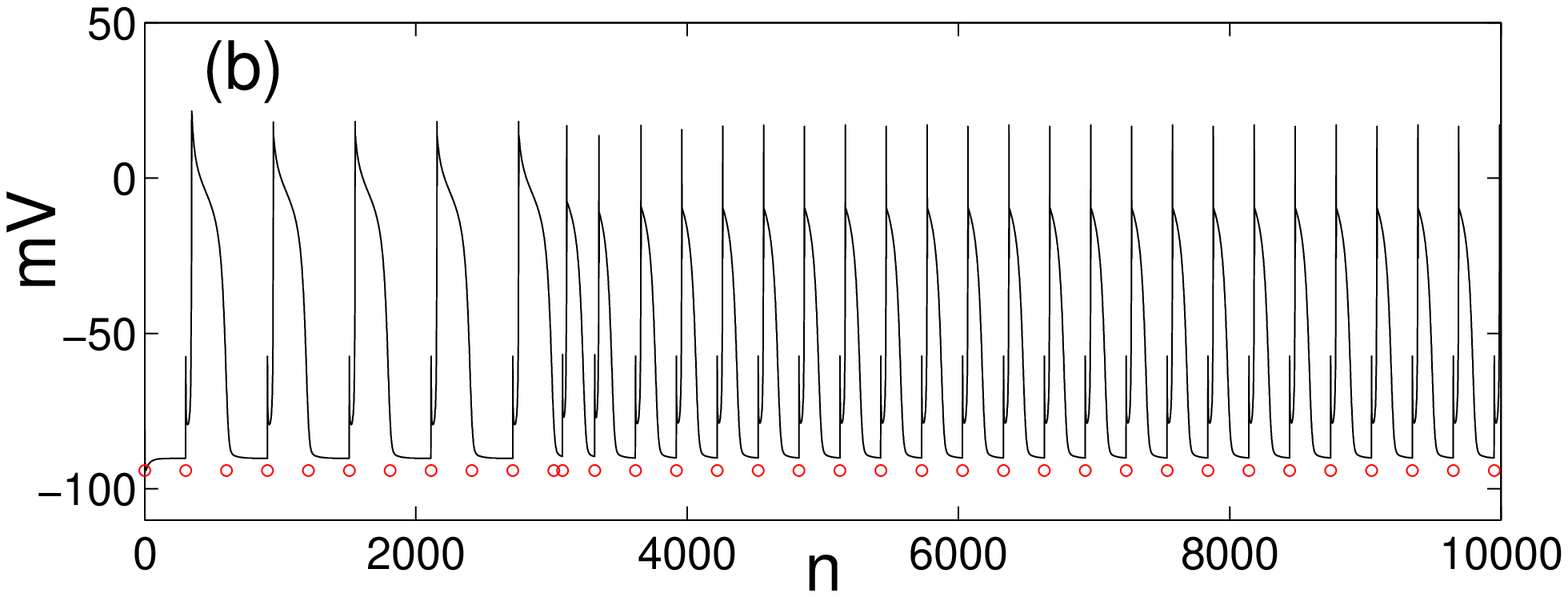}
\includegraphics[scale=0.4]{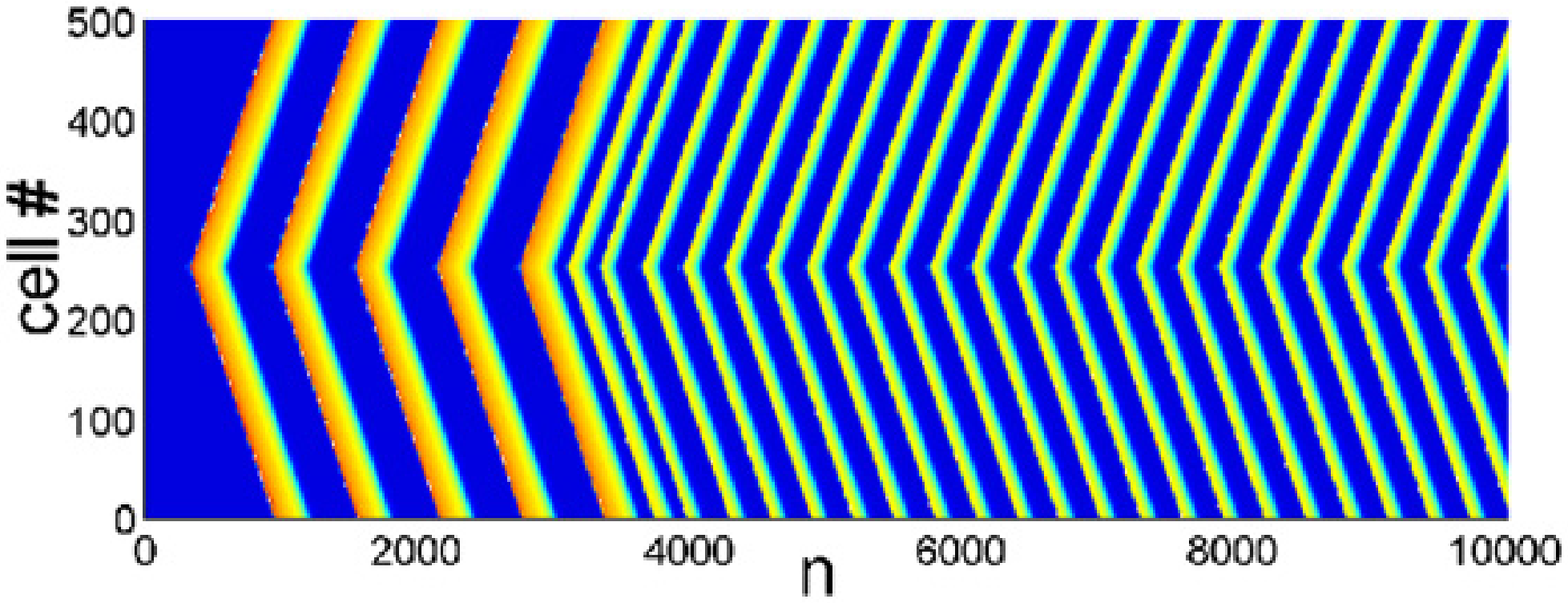}
\caption{(Color online) Waveforms of $V_{n,250}$ triggered by the
stimuli (red circles) and images of waves in the circle of 500
electrically coupled cells. Regime of frequency ratio 1:2 - (a) and
a change to the regime of 1:1 caused by the proper position of the
additional trigger pulse - (b)}\label{Images1}
\end{figure}

\begin{figure}[t]
\includegraphics[scale=0.4]{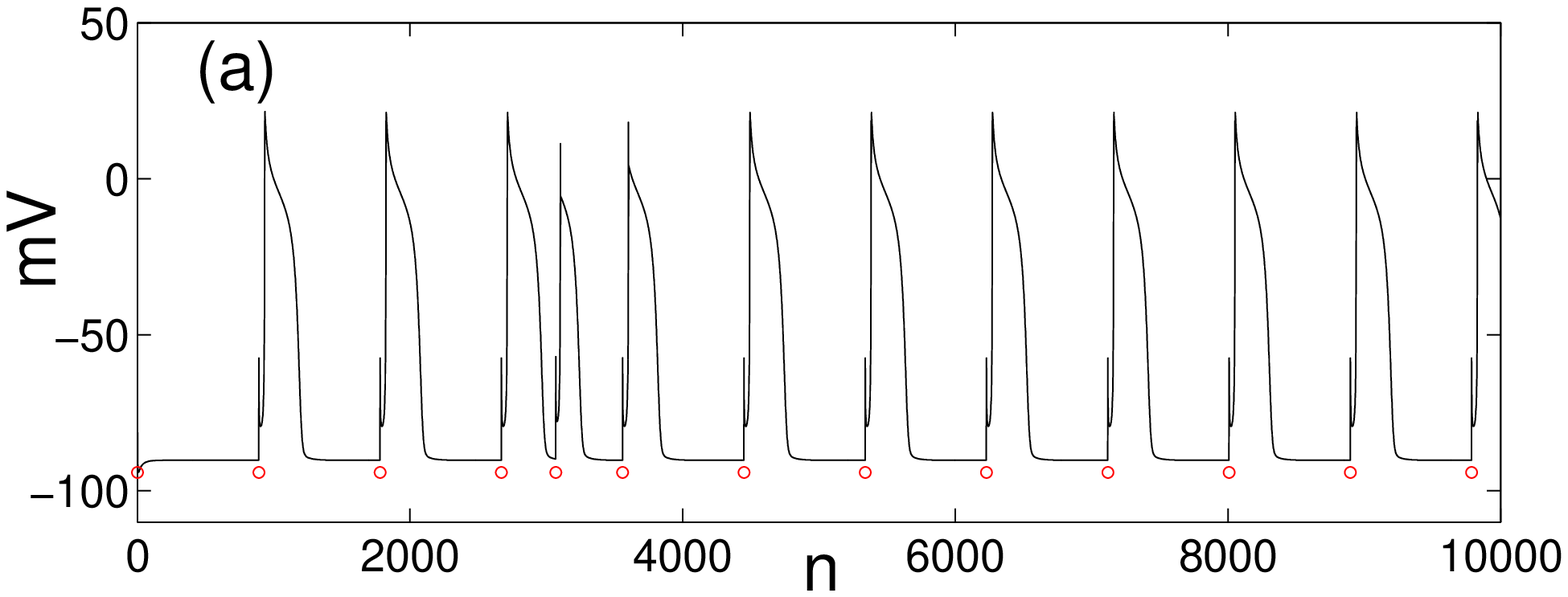}
\includegraphics[scale=0.4]{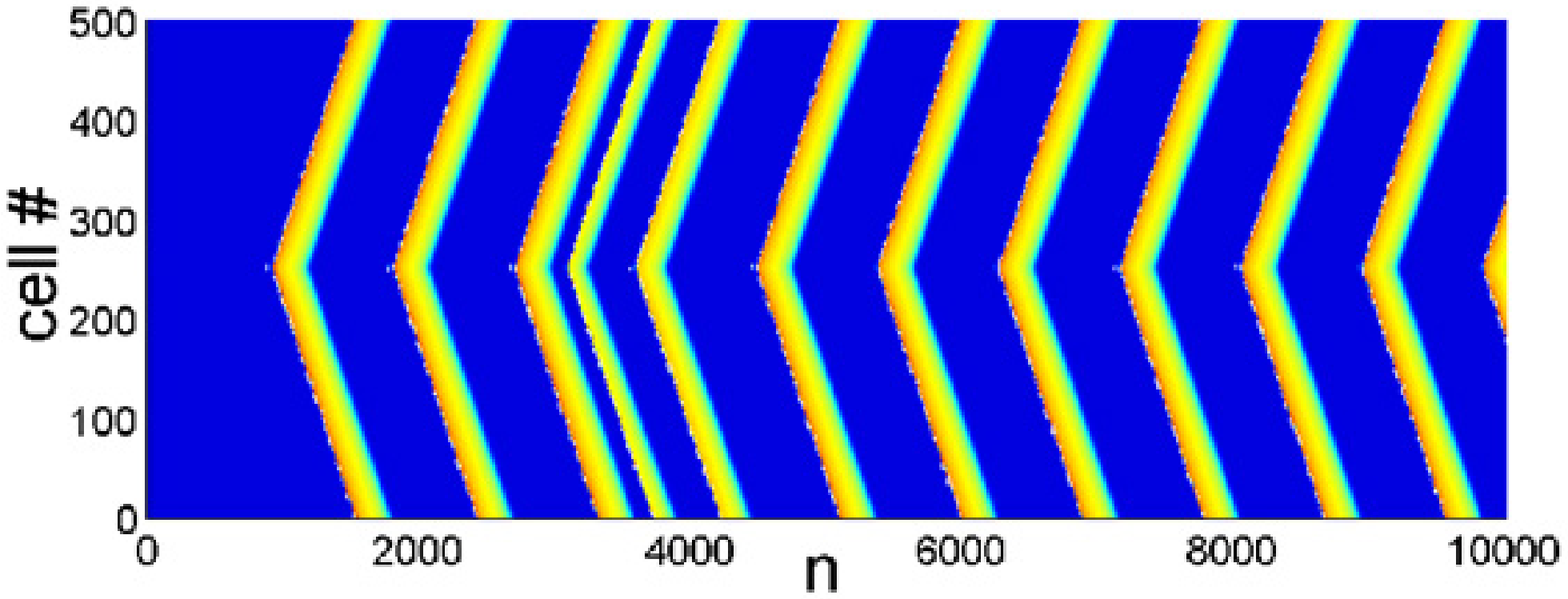}
\includegraphics[scale=0.4]{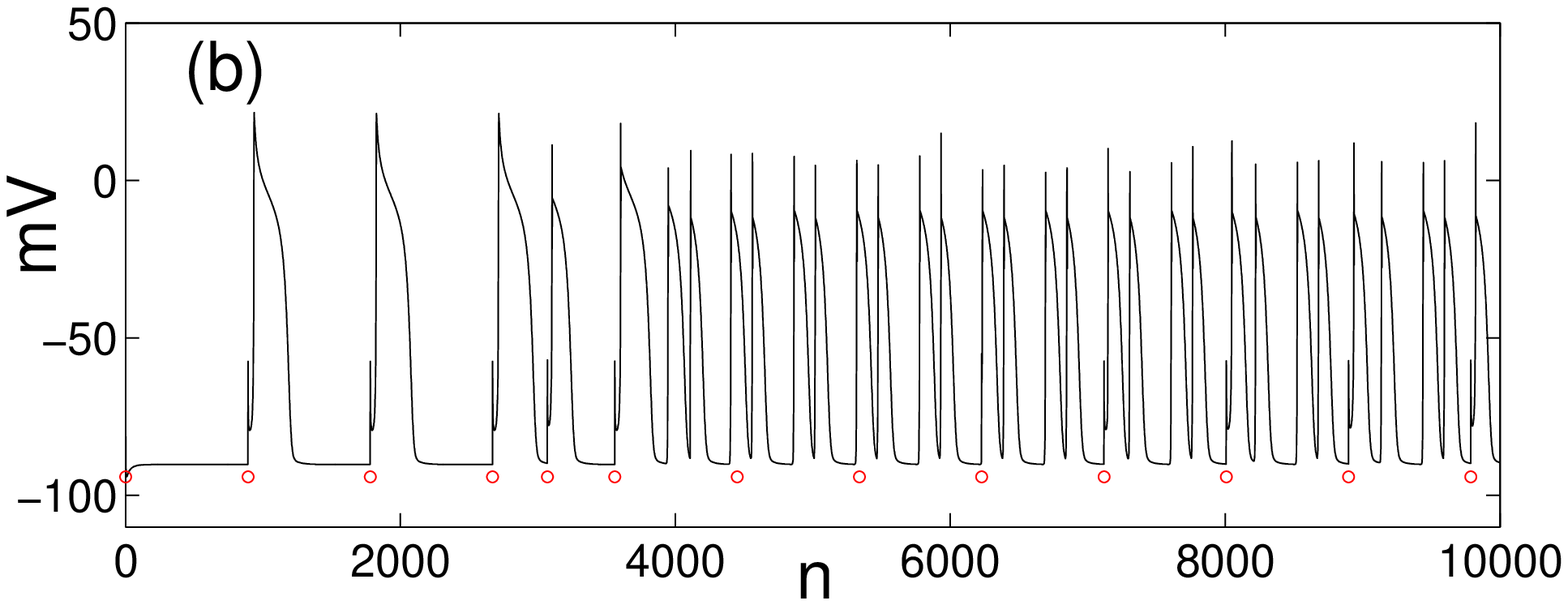}
\includegraphics[scale=0.4]{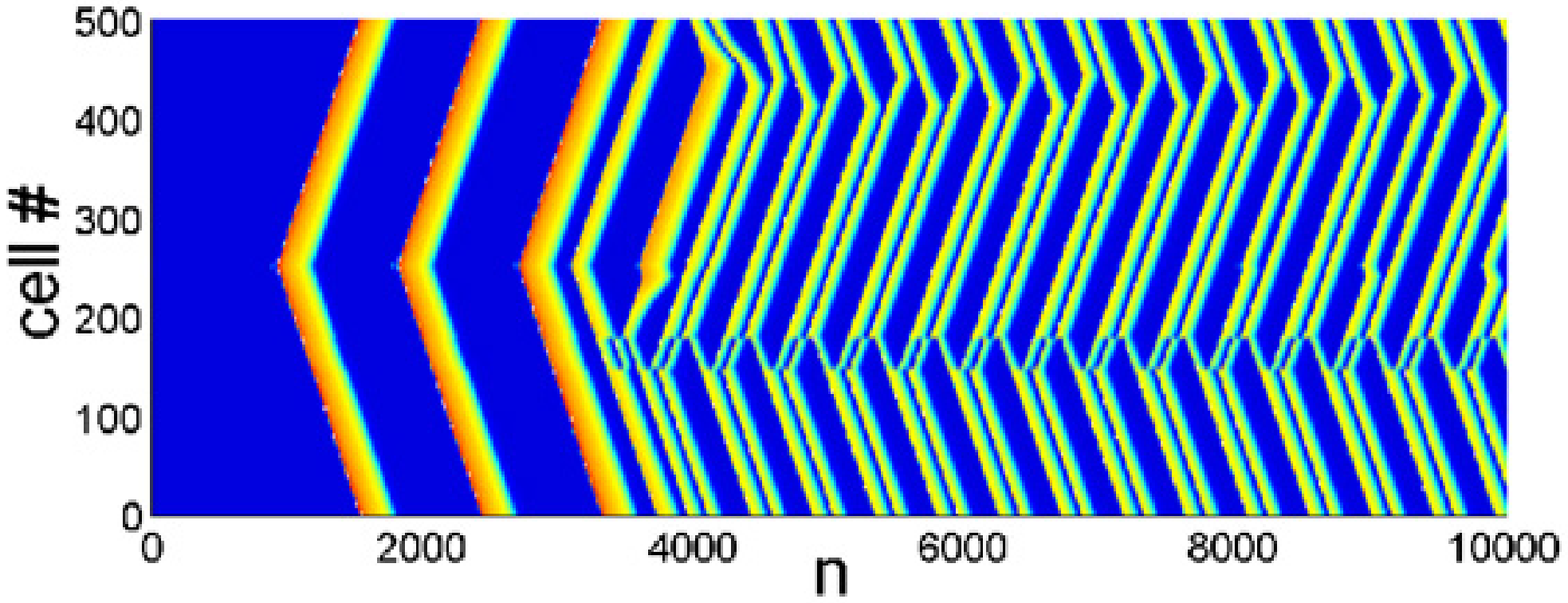}
\caption{(Color online) Waveforms of $V_{n,250}$ triggered by the
stimuli (red circles) and images of waves in the circle of 500
electrically coupled cells, without (a) and with (b) a group of
"damaged" cells.}\label{Images2}
\end{figure}

To illustrate some of the effects captured by such a coupling model
we consider a one-dimensional chain with periodic boundary
conditions which contains 500 cells coupled to nearest neighbors.
Electrical activity of this circle of cells was initiated by
excitation of a single cell (cell number 250) using a periodic
sequence of triggering pulses and one additional pulse, whose
position in time was controlled independently of the periodic
sequence. The amplitude of the trigger pulses was set $A=0.4$ and
the duration of each pulse was one iteration long. The parameters of
the cell model were selected as $\alpha=3.2$, $\beta_x=-2.5780$,
$x_p=-0.8$, $y_s=1.3$, $\mu=0.001$, $g=0.1$, $y_{th}=0.01$,
$S_s=0.399$, $S_a=0.4$ and $g_s=0.03$. The parameters of the gap
junction model were set equal to $g_{gap}=0.004$ and
$\mu_{gap}=0.0001$. The results of simulations are shown is
Figs.~\ref{Images1} and \ref{Images2}.

Figure ~\ref{Images1} illustrates the case of the coexistence of two
regimes of wave activity with the frequency ratios 1:2 and 1:1. In
Fig.~\ref{Images1}(a) the additional trigger pulse has a phase that
inserts perturbations into the wave behavior, but does not change
the regime of activity after the transient is terminated. When the
position of the additional pulse was moved closer to the end of the
previous AP it was sufficient to switch the regime of activity from
1:2 to 1:1 frequency ratio. This effect is similar to the one shown
in Fig.~\ref{wfpAP}, but here the memory effects were controlled
directly with (\ref{Sn}). The sudden change of frequency ratio of
the waves is typical for cases of high frequency triggering pulses
and was demonstrated before in simulations with more realistic
conductance-based models, see for example~\cite{Stamp02,Kanakov07}.

Figure ~\ref{Images2} shows the effect that can be caused by
existence of a small group of "damaged" cells on the wave activity
of the chain. In this case the period between the triggering
impulses is larger than the APD of normal cells and, when the
parameters of all cells are the same, the APs with 1:1 frequency
ratio is the only stable regime of wave activity in the chain. It is
illustrated in Fig.~\ref{Images2}(a) where the periodic waves
perturbed by the additional trigger pulse recover very fast. The
situation with the recovery process changed dramatically after a
group of 30 cells (with indexes $i$ from 150 to 180) was "damaged".
The "damage" was done by shortening their APD by using $y_s=1.2$
instead of $y_s=1.3$. It is interesting that this group of cell did
not show signs of "bad' behavior during the periodic sequence of
triggering pulses, see Fig.~\ref{Images2}(b). Indeed the three waves
at the beginning are almost identical to the waves shown in
Fig.~\ref{Images2}(a). However after the additional trigger pulse
perturbed the wave activity of the chain the regime of 1:1
oscillations did not recover and the system switched to a new high
frequency wave pattern. This new pattern is characterized by
doublets of AP waves. The group of "damaged" cells formes a new
source of wave excitation, see Fig.~\ref{Images2}(b) cells from
$i$=150 to $i$=180.

\section{Conclusion}\label{sec5}

We have considered basic elements of a map-based approach to the
design of a simple, computationally efficient model for replication
of action potential in a non-pacemaker cardiac cell. This paper is
focused mainly on the basic methods of discrete-time dynamics for
the model design, rather than on fitting its parameters to capture
the characteristics of a specific cardiac cell. The map-based models
tuned for replicating the dynamics of specific cardiac cells will be
published elsewhere.

\section{Acknowledgement}\label{sec6}

The author thanks G.V. Osipov, O.I. Kanakov and A.M. Hunt for a
useful discussion and drawing the author's attention to the interest
in application of map-based approaches for modeling of non-pacemaker
cardiac cells.

\end{document}